\documentclass[PRD,twocolumn,showpacs,preprintnumbers,nofootinbib,amsmath,amssymb]{revtex4}
\usepackage[dvips]{graphics}
\usepackage{graphicx}
\usepackage{dcolumn}
\usepackage{bm}
\usepackage{amssymb,amsmath}
\usepackage{epsfig}
\usepackage{color}
\usepackage{amsfonts}
\usepackage{amscd}
\usepackage{dsfont}

\newcommand{\beq}{\begin{equation}}
\newcommand{\eeq}{\end{equation}}
\newcommand{\bga}{\begin{gathered}}
\newcommand{\ega}{\end{gathered}}
\newcommand{\beqa}{\begin{eqnarray}}
\newcommand{\eeqa}{\end{eqnarray}}

\begin{document}

\title{Testing Parity-Violating Mechanisms with Cosmic
Microwave Background Experiments}

\author{Vera Gluscevic and Marc Kamionkowski}
\affiliation{California Institute of Technology, Mail Code 350-17,
Pasadena, CA 91125\\}

\date{\today}

\begin{abstract}
Chiral gravity and cosmological birefringence both provide
physical mechanisms to produce parity-violating TB and EB
correlations in the cosmic microwave background (CMB)
temperature/polarization.  Here, we study how well these two
mechanisms can be distinguished if non-zero TB/EB correlations
are found.  To do so, we evaluate the correlation
matrix, including new TB-EB covariances.  We find that the
effects of these two mechanisms on the CMB are highly
orthogonal, and can thus be distinguished fairly well in case of
a high--signal-to-noise detection of TB/EB correlations.  An
Appendix evaluates the relative sensitivities of the BB, TB, and
EB signals for detecting a chiral gravitational-wave background.
\end{abstract}


\maketitle
\section{Introduction}\label{intro}

Both inflation \cite{Guth:1980zm} and late-time cosmic
acceleration \cite{Riess:1998cb} require new physics beyond
general relativity and the standard model (SM) of particle physics.
Since the SM violates parity (P) within the weak sector and is
presumably only a low-energy limit of a grand unified theory, it
is natural to inquire whether there are manifestations of P
violation in the new physics responsible for cosmic inflation
and/or late-time acceleration.

For example, a coupling of the quintessence field to the
pseudo-scalar of electromagnetism would manifest itself as
cosmological birefringence (CB) \cite{Carroll:1998zi}, a
rotation of the linear polarization of electromagnetic waves as
they propagate through the Universe.  Parity violation has been
introduced in inflation through modifications of gravity that
produce a difference in the amplitude of right (R) and
left (L) circularly polarized gravitational waves (GWs) in the
inflationary GW background.  These include
the addition of Chern-Simons terms to the Einstein-Hilbert action 
\cite{Lue:1998mq}; chiral gravity, wherein there is a
different Newton's constant for R and L gravitational waves
\cite{Contaldi:2008yz}; and gravity
at a Lifshitz point \cite{Takahashi:2009wc}.  We refer
collectively to these inflationary mechanisms as chiral gravity.

Since the CMB polarization can be decomposed into two modes of
opposite parity---E modes, or the gradient part, and B modes, or
the curl part \cite{Kamionkowski:1996ks,Zaldarriaga:1996}---a
cross-correlation between the E and B modes would, if detected,
be a sign of parity violation \cite{Lue:1998mq}; and similarly
for a correlation between the temperature (T) and the B mode.
Chiral GWs induce TB/EB correlations at the CMB last scattering
surface (LSS) \cite{Lue:1998mq,Contaldi:2008yz}, while CB
induces P violation by rotating the primordial polarization
afterwards \cite{Lue:1998mq,Lepora:1998ix}. 

An early analysis of CMB data suggested a possible CB with
rotation angle $\sim6^\circ$ \cite{Feng:2006dp}, but current
constraints are less than a few degrees
\cite{constraints,Komatsu}.
 Ref.~\cite{Saito:2007kt} showed that
WMAP does not have enough sensitivity to test chiral gravity and
discussed prospects for detection of chiral GWs with
Planck and CMBPol.

In this paper, we quantify how well the effects of CB and chiral
gravity can be distinguished, in case of a
positive detection of EB/TB correlations.  We find that the
effects of these two mechanisms are orthogonal to a very high
degree, and we show that the earlier tentative
detections of CB, if true, could not have been attributed to
chiral gravity.  We perform these forecasts for WMAP \cite{WMAP}, SPIDER \cite{Crill:2008rd}, Planck \cite{Planck_0}, CMBPol (EPIC) \cite{CMBPol}, and a cosmic-variance--limited experiment.

The plan of this paper is as follows:  In \S\ref{chirality}, we
forecast the sensitivity of CMB experiments to gravitational
chirality and in \S\ref{birefringence} to CB.  \S\ref{chi_alpha}
calculates how well the two effects can be distinguished, and in
\S\ref{conclude} we make concluding remarks. In Appendix A, we derive the elements of the power-spectra covariance matrix, and in Appendix B, we
evaluate the relative sensitivities of the BB,
TB, and EB signals to a chiral gravitational-wave background,
finding that the best sensitivity comes from the BB signal, in
disagreement with an earlier claim \cite{Contaldi:2008yz}.

\section{Constraining Gravitational Chirality}\label{chirality}

\subsection{Effects of Gravitational Chirality on the CMB Polarization}\label{chirality_fls}
 
If linearized gravity prefers one handedness (i.e., if it is
chiral), then the power spectra of the L and R GWs may have
different amplitudes and thus induce non-vanishing TB and EB
correlations at the LSS
\cite{Lue:1998mq,Contaldi:2008yz,Takahashi:2009wc}. Measurements
of these correlations can provide an estimate of the chiral
asymmetry with a variance due to the finite precision of the
instrument and cosmic variance (CV).

We first want to quantify the chirality by introducing an
appropriate chirality parameter and show how the CMB
polarization map depends on this parameter. To have B modes at
the LSS, we need primordial GWs, or in other words, a non-zero
tensor-to-scalar ratio,
\beq
\bga
     r \equiv A_t/A_s,\hspace{0.5cm} A_t=\frac{r}{1+r},
\ega
\label{tensor_to_scalar_ratio}
\eeq
where $A_t$ and $A_s$ are, respectively, the fractional
contributions of tensor and scalar modes to the
TT quadrupole. Each
one of the six CMB temperature/polarization power spectra---TT,
EE, BB, TE, TB, and EB---have a tensor component proportional to
$A_t$, while TT, EE, and TE additionally have a scalar component
proportional to $A_s$. The tensor-to-scalar ratio $r$ is
currently constrained to be $\leq 0.22$ at a $95\%$ confidence
level \cite{Komatsu}.

The TB and EB power spectra are proportional to the difference of
the L- and R-mode contributions to the GW (tensor) power spectra, $P^{t,L}(k)$ and $P^{t,R}(k)$.
These P-violating power spectra are \cite{Saito:2007kt},
\beq
     {C^{XX'}_l} = (4\pi)^2 \int {k^2dk}[P^{t,L}(k)-P^{t,R}(k)]
     \Delta _l^{X}(k)\Delta _l^{X'}(k),
\label{p_violating_correlations_integral}
\eeq
while the tensor part of the P-conserving correlations are
\beq
     {C^{XX'}_l} = (4\pi)^2 \int
     {k^2dk}[P^{t,L}(k)+P^{t,R}(k)]\Delta _l^{X}(k)\Delta
     _l^{X'}(k),
\label{p_conserving_correlations_integral}
\eeq
where $P^{t,L}(k)$ and $P^{t,R}(k)$ are the L- and R-mode power
spectra, $\Delta _l^{X}(k)$ is the radiation transfer function
for $X$, and $X,X'=\{T, E, B\}$. 
\begin{figure}[htbp]
\begin{center}
\includegraphics[height=8cm,keepaspectratio=true]{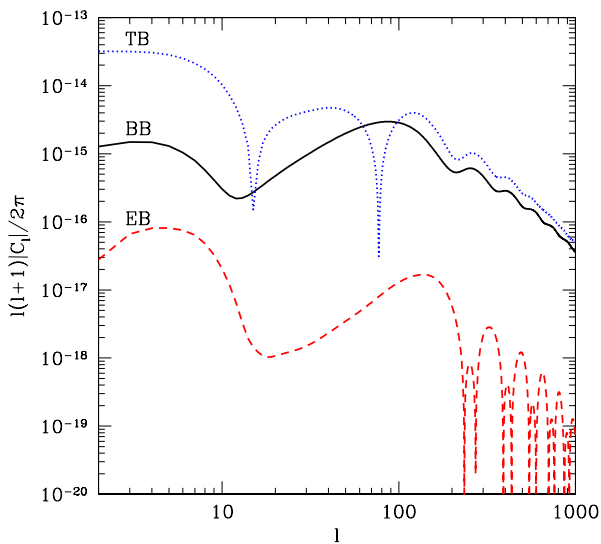}
\caption{B-mode power spectra for $r=0.22$ and $\Delta \chi =0.2$.\label{ps}}%
\end{center}
\end{figure}
Following Ref.~\cite{Saito:2007kt}, we define a chirality
parameter $\Delta \chi$ as
\beq
\bga
     P^{t,L}(k)\equiv \frac{1}{2}(1+\Delta \chi)P^t(k),\\
     P^{t,R}(k)\equiv \frac{1}{2}(1-\Delta \chi)P^t(k),
\ega
\label{chi_definition}
\eeq
where
\beq
     P^{t}(k)\equiv P^{t,L}+P^{t,R}.
\label{Pt}
\eeq
Maximal P violation occurs when there are GWs of only one
handedness; $\Delta \chi=1$ corresponds to fully left-handed,
and $\Delta \chi=-1$ to fully right-handed GWs.  To illustrate,
we show BB, TB, and EB power spectra for $r=0.22$ and $\Delta
\chi =0.2$ in Fig.~\ref{ps}.

To calculate the uncertainty with which $\Delta \chi$ can be
estimated with different experiments we use a Fisher-matrix
analysis \cite{Jungman:1995bz}, employing the null hypothesis,
$C_l^{EB}=C_l^{TB}=0$.  This ensures that the TB and EB power
spectra do not have cross correlations with the other four power
spectra. The reciprocal value of the variance $\sigma_{\Delta
\chi}^2$ is then given by \cite{Kamionkowski:1996ks}
\beq
     \sigma _{\Delta \chi}^{ - 2} = 
     \sum\limits_l {\sum\limits_{A, A'} {} } \frac{{\partial
     C_l^A}} {{\partial \Delta \chi}}\frac{{\partial C_l^{A'}}}
     {{\partial \Delta \chi}}{[{\Xi_{l}}^{ - 1}]_{AA'}},
\label{sigma_matrix_chi}
\eeq
where $A, A' = \{TB, EB\}$, and ${\Xi_{l}}$ is the TB-EB part of
the power-spectrum covariance matrix;\footnote{Under the
null hypothesis $\Delta\chi=0$, the $2\times 2$ TB-EB part of the inverted
$6\times 6$ covariance matrix is the same as the inverse of the
$2\times 2$ TB-EB matrix.}  This covariance matrix is derived in Appendix A.
The partial derivatives in
Eq.~(\ref{sigma_matrix_chi}) can be evaluated by noting from
Eqs.~(\ref{p_violating_correlations_integral}) and
(\ref{chi_definition}) that 
\beq
(\partial C_l^{TB/EB}/{\partial \Delta \chi}) = C_l^{TB/EB}(\Delta \chi = 1).
\eeq
We obtain the TB/EB power spectra by modifying CMBFAST
\cite{Seljak:1996is} and using a $\Lambda$CDM model consistent with
WMAP-5 \cite{Komatsu} parameters.  

\subsection{Numerical Results: Forecasts for Errors to $\Delta
\chi$}\label{chirality_num}

\begin{center}
\begin{table}[htbp]
    \begin{tabular}{ | l | l | l | l | p{5cm} |}
    \hline
    Instrument & $\theta _{\text{fwhm}}$ [arcmin] & NET $[\mu
\text{K}\sqrt{\text{sec}}]$ &  $t_{\text{obs}}$ [years] \\ \hline
    WMAP-5 & 21 & 650 & 5 \\ \hline
    SPIDER & 60 & 3.1 & 0.016 \\ \hline
	 Planck & 7.1 & 62 & 1.2 \\ \hline
    CMBPol & 5 & 2.8 & 4 \\ \hline
	 CV-limited & 5 & 0 & 1.2 \\
    \hline
    \end{tabular}
\caption{Instrumental parameters from
     Ref.~\cite{Crill:2008rd,Pullen:2007tu,Planck,Bock:2008ww}
     for the five experiments considered in this paper.  The
     parameters are the beamwidth $\theta _{\text{fwhm}}$,
     noise-equivalent temperature NET, and observation time
     $t_{\text{obs}}$.\label{instruments}}
\end{table}
\end{center}

We now forecast the sensitivities to chiral
GWs of the following five experiments: (i) WMAP-5, (ii) SPIDER's 150 GHz
channel, (iii) Planck's 143 GHz channel, (iv) CMBPol's (EPIC-2m)
150 GHz channel, and (v) a CV-limited experiment. The
corresponding instrumental parameters are given in Table
\ref{instruments}. Note that the noise-equivalent temperature
NET is related to the temperature/polarization pixel-noise
variances, $\sigma _{T/P}$, as 
$\sigma_T^2/N_{\text{pix}}= (\text{NET})^2/t_{\text{obs}}$,
where $\sigma_P=\sqrt{2}\sigma_T$. We take $f^0_{\text{sky}}=1.0$ (the fraction of the sky surveyed), and
$f_{\text{sky}}=0.7$ (the fraction of the sky used in the analysis), for all experiments, except for
SPIDER, where $f^0_{\text{sky}}=f_{\text{sky}}=0.5$.

\begin{figure}[htbp]
\begin{center}
\includegraphics[height=8cm,keepaspectratio=true]{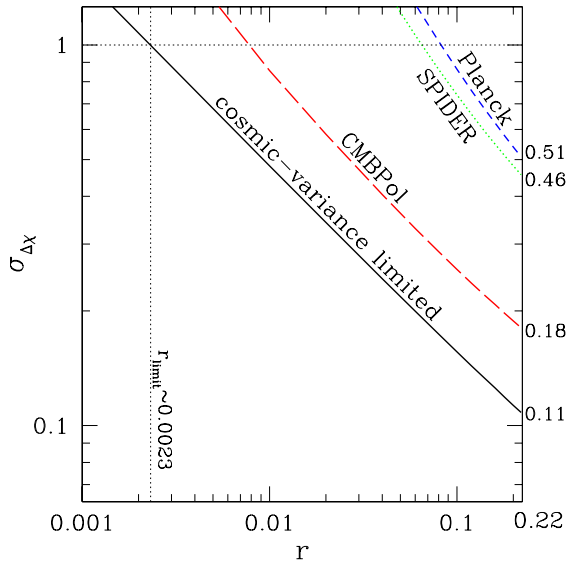}
\caption{1$\sigma$ error on the gravitational chirality parameter $\Delta \chi$, for five different CMB experiments, for the fiducial value of $\Delta \chi=0$. The horizontal dotted line is at $\sigma _{\Delta \chi}=1$ and represents maximal P violation. In the region above this line, the chirality is non-detectable. The WMAP-5 curve lies entirely above the non-detection line.\label{deltax}}%
\end{center}
\end{figure}
Fig.~\ref{deltax} shows the 1$\sigma$ error of the
estimate of $\Delta \chi$ as a function of tensor-to-scalar
ratio $r$. The error increases with decreasing $r$, which
implies the existence of a critical value of $r$ below which a
1$\sigma$-level detection becomes impossible even for maximal P
violation (when $\sigma _{\Delta \chi}\geq 1$). This value is
far above the current upper limit for WMAP-5 (compare to
Ref.~\cite{Saito:2007kt}), and so WMAP-5 can give no constraints
on chiral gravity. Prospects are more optimistic for the
next-generation CMB data releases. The critical $r$ is about $0.064$ for SPIDER,
$0.082$ for Planck, $0.0079$ for CMBPol, and $0.0023$ for the
CV-limited experiment. If $r$ is just below the
current detection limit of $0.22$ \cite{Komatsu}, $\Delta \chi$
will be detectable at the 1$\sigma$ level if it is greater than
$0.46$, $0.51$, $0.18$, and $0.11$ for these four instruments,
respectively. If we consider the 3$\sigma$ confidence level,
the corresponding minimum detectable values are larger by a
factor of $\sim3$.

\begin{figure}[htbp]
\begin{center}
\includegraphics[width=11cm,keepaspectratio=true]{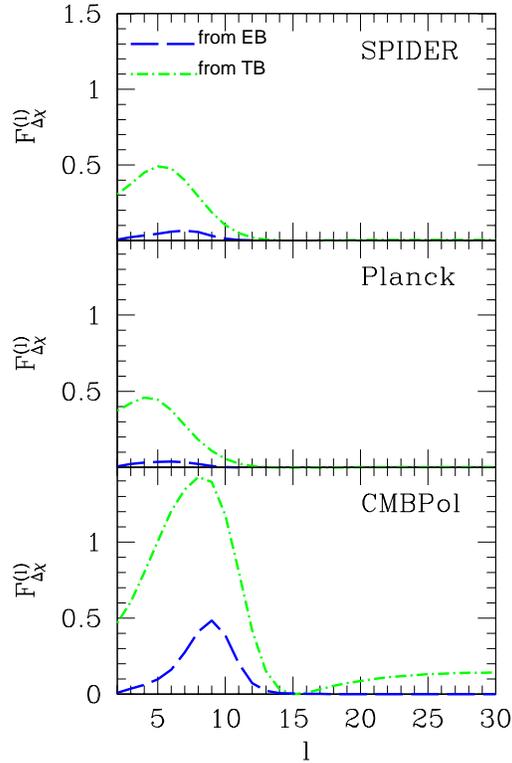}
\caption{Diagonal (TB,TB and EB,EB) summands of
     Eq.~(\ref{sigma_matrix_chi}), for $r=0.22$, are plotted against the
     multipole $l$ to show that the constraint
     to $\Delta \chi$ comes primarily from the TB power spectrum
     at $l\sim 7$.\label{x_contributions}}%
\end{center}
\end{figure}

To conclude this Section, we show how different multipoles $l$
contribute to the sum of Eq.~(\ref{sigma_matrix_chi}),
separating the contribution from TB and EB, in
Fig.~\ref{x_contributions}. In this plot, only the TB/EB
summands of Eq.~(\ref{sigma_matrix_chi}) are plotted against
$l$, for $r=0.22$, for SPIDER, Planck, and CMBPol. The
off-diagonal terms that contain the covariance between TB and EB
are negligible. The major contribution to $\sigma_{\Delta
\chi}^{-2}$ for all five experiments comes from the TB power
spectrum, from low multipoles, $l\sim 7$. Thus, large
angular scales in TB (at $l\leq 10$) contain most of
the information about gravitational chirality.

\section{Constraining Cosmological Birefringence}\label{birefringence}

Cosmological birefringence rotates the linear polarization at
each point on the sky by an angle $\Delta\alpha$, and this
rotation induces TB/EB power spectra
\beq
     C^{TB,\text{rot}}_l = 2\Delta \alpha C^{TE}_l, \qquad
     C^{EB,\text{rot}}_l = 2\Delta \alpha C^{EE}_l.
\label{C_rot}
\eeq
The error $\sigma _{\Delta \alpha}$ to which $\Delta\alpha$ can
be measured is given by
\beq
     \sigma _{\Delta \alpha}^{ - 2} =\sum\limits_l
     {\sum\limits_{A, A'} {} }
     \frac{{\partial C_l^A}}
     {{\partial \Delta \alpha}}\frac{{\partial C_l^{A'}}}
     {{\partial \Delta \alpha}}{[{\Xi_{l}}^{ - 1}]_{AA'}}.
\label{sigma_matrix_alpha}
\eeq

Using the same instrumental parameters as in \S\ref{chirality_num}, and for $r=0.22$, we obtain the following
1$\sigma$ errors for the CB rotation angle: from WMAP-5,
$3.2^{\circ}$; from SPIDER, $0.9^{\circ}$; from Planck, $15.9'$;
from CMBPol, $9.4"$; and from a CV-limited experiment, 1.9
$\mu$arcsec, in good agreement with previous forecasts
\cite{Lue:1998mq,Kamionkowski:2008fp,Yadav:2009eb,Gluscevic:2009mm}.

\begin{figure}[htbp]
\begin{center}
\includegraphics[height=11cm,keepaspectratio=true]{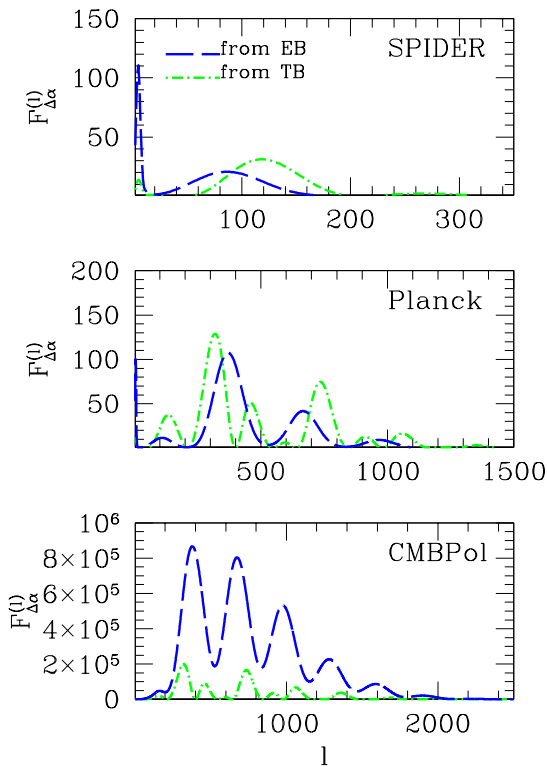}
\caption{Diagonal (TB,TB and EB,EB) summands of
     Eq.~(\ref{sigma_matrix_alpha}), for $r=0.22$, are plotted against the
     multipole $l$ to show that the constraints to $\Delta
     \alpha$ from future CMB experiments will come primarily
     from $l$'s of $\sim$100, 500, or 700 (depending on the
     instrument).\label{alpha_contributions}}%
\end{center}
\end{figure}

In Fig.~\ref{alpha_contributions}, we plot, separately, the
contributions from only TB and only EB correlation to the sum in
Eq.~(\ref{sigma_matrix_alpha}), as a function of multipole
moment $l$, for the cases of SPIDER, Planck, and CMBPol, for
$r=0.22$. The off-diagonal terms that contain the covariance
between TB and EB are small. The dominant contribution to the
constraint on $\Delta \alpha$ comes from the TB correlation for
WMAP-5, and from EB for the higher-precision
instruments. Different multipoles give the leading summands in
$\sigma_{\Delta \alpha}^{-2}$ for different instruments, but
unlike the case of GW chirality, small angular scales
($l\gtrsim100$) always dominate the sum. 

\section{Separating Gravitational Chirality from Cosmological Birefringence}\label{chi_alpha}

In this Section, we ask how well the effects of chiral
gravity and CB can be distinguished, assuming that a TB/EB
correlation has been detected.

\subsection{First-Order Effects on the EB and TB Correlations}\label{chi_alpha_fls}

To first order in $\Delta \alpha$ and $\Delta \chi$, the TB/EB
power spectra are a sum of a part $C_l^{A,\text{chi}}$ due to chiral GWs
and a part $C_l^{A,\text{rot}}$ due to CB.  The combined EB and
TB power spectra can be written,
\beq
\bga
     C^{TB,\text{obs}}_l = \Delta \chi C_l^{TB,t}(\Delta \chi =
     1) + 2\Delta \alpha C^{TE}_l,\\
     C^{EB,\text{obs}}_l = \Delta \chi C_l^{EB,t}(\Delta \chi =
     1) + 2\Delta \alpha C^{EE}_l,
\ega
\label{C_obs_all}
\eeq
where the superscript $t$ indicates the tensor-induced part of the
power spectrum, while the absence of it denotes the full power
spectrum, including the scalar part.

\begin{figure}[htbp]
\begin{center}
\includegraphics[height=11cm,keepaspectratio=true]{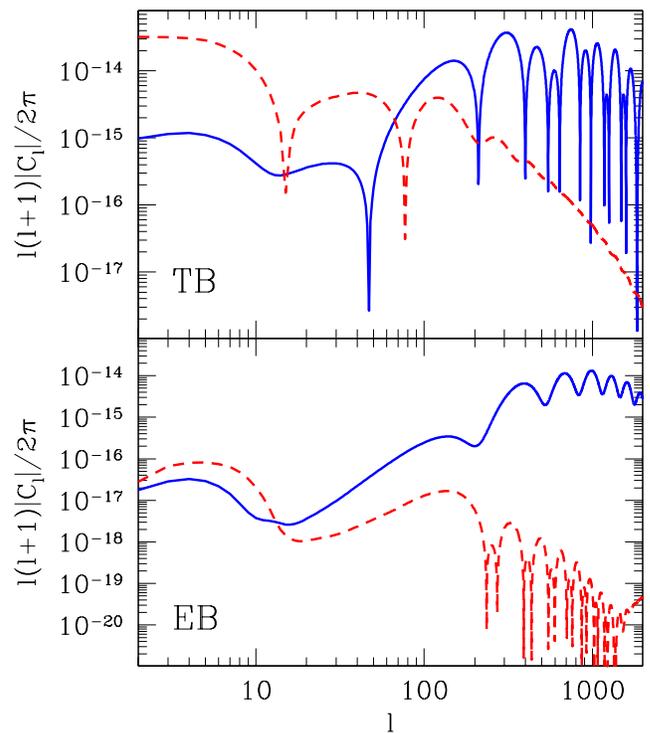}
\caption{We show TB and EB power spectra from chiral GWs for
     $\Delta \chi = 0.2$ and $r=0.22$ (dashed red curves) and from
     cosmological birefringence for $\Delta \alpha=5'$ (solid
     blue curves).
\label{combined}}%
\end{center}
\end{figure}
Fig.~\ref{combined}, which shows $C^{A,\text{chi}}_l$ and
$C^{A,\text{rot}}_l$, demonstrates that the contributions from these
two mechanisms are qualitatively different. Our goal now is to
quantify how well they can be distinguished, given the finite
precision of the temperature/polarization maps.

The Fisher matrix for $\Delta \alpha$ and $\Delta \chi$ has the
following entries (see \cite{Kamionkowski:1996ks}),
\beq
\mathcal{F}_{ij} = \sum\limits_l {\sum\limits_{A,A'} {} }
\frac{{\partial C_l^A}} {{\partial {a_i}}}\frac{{\partial
C_l^{A'}}} {{\partial {a_j}}}{[{\Xi_l ^{ - 1}}]_{AA'}},
\label{fisher_all}
\eeq
where $i,j=\{1,2\}$; $a_{i}$ and $a_{j}$ are the elements of
$\vec a=(\Delta \alpha, \Delta \chi)$; $A,A'=\{TB, EB\}$; and
$\mathcal{F}$ is the inverse of the covariance matrix between
$\Delta \alpha$ and $\Delta \chi$. The derivatives in
Eq.~(\ref{fisher_all}) can be calculated using
Eq.~(\ref{C_obs_all}), and ${[{\Xi ^{ - 1}}]_{AA'}}$ is the
inverse of the TB-EB covariance matrix given by
Eq.~(\ref{covariance_matrix_entries}) of Appendix A. Once again, we employ the
null hypothesis\footnote{Even in the case where we work around
non-zero fiducial values, the effect of the off-diagonal terms is negligible and the
covariance matrix can be treated as a block-diagonal matrix to
good precision. In addition, the cross-terms between TB/EB
and the other four power spectra in Eq.~(\ref{fisher_all})
vanish, to the first order in small parameters, so we really
only need to consider TB and EB.} $\Delta \alpha=\Delta
\chi=0$.

\subsection{Numerical Results: Constraints on the $\Delta \alpha$-$\Delta \chi$ parameter space}\label{chi_alpha_num}

\begin{figure}[htbp]
\begin{center}
\includegraphics[height=11cm,keepaspectratio=true]{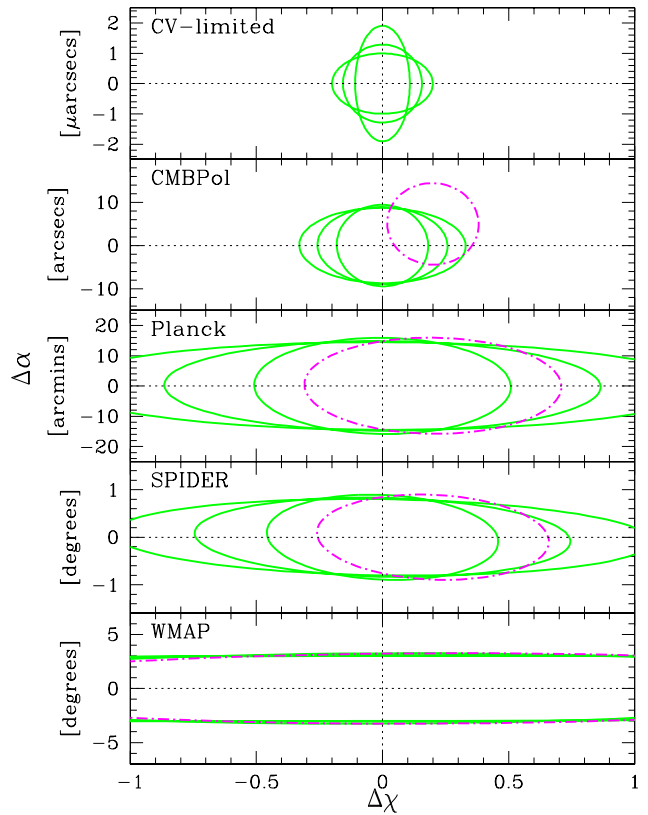}
\caption{Constraints on the allowed $\Delta \alpha$-$\Delta
     \chi$ parameter space are shown for the case of null
     detection with different instruments. The solid-line
     ellipses on each plot are for the fiducial value of zero
     for both parameters, and for the following values of
     tensor-to-scalar ratios (going from the narrowest to the widest ellipse in the $\Delta \chi$ direction): $0.22$, $0.1$, and $0.06$. The
     dot-dashed ellipse in each plot is for $r=0.22$, for a
     model with $\Delta \chi=0.2$ and $\Delta \alpha=5"$. In
     both models, the tilt of the ellipses is almost negligible,
     which means that the two P-violating mechanisms are
     separable to high accuracy, provided there is detection
     with high statistical significance. Thus, the constraints
     on both parameters are almost the same as those
     calculated in the previous Sections.\label{ellipses}}%
\end{center}
\end{figure}

Fig.~\ref{ellipses} shows 1$\sigma$ error ellipses in the
$\Delta \alpha$-$\Delta \chi$ parameter space, for the null
hypothesis $\Delta\alpha=\Delta\chi=0$,
with WMAP-5, SPIDER, Planck, and CMBPol, for a
range of tensor-to-scalar ratios. In addition, each plot shows a
1$\sigma$-error ellipse for a different set of fiducial values:
$\Delta \chi=0.2$ and $\Delta \alpha=5"$.  The
ellipses for this model are merely shifted in the
$\Delta \alpha$-$\Delta \chi$ space, but are otherwise not
significantly different from the null-hypothesis ellipses.
From Fig.~\ref{ellipses}, we see that once we take into account
the covariance between $\Delta \alpha$ and $\Delta \chi$, the
results differ very slightly from the two cases where we had
only one of the P-violating mechanisms acting on the CMB. This
is clear from the fact that the ellipses show
very little tilt in $\Delta \alpha$-$\Delta \chi$ space. We
conclude that if non-vanishing TB/EB correlations are detected
with high statistical significance, we will be able to
distinguish CB from gravitational chirality to a high degree. 

This result can also be explained from
Figs.~\ref{x_contributions} and \ref{alpha_contributions}, which
show that the $\Delta \chi$ constraint comes primarily from TB
at low $l$'s, while the $\Delta \alpha$ constraint comes
primarily from EB at high $l$'s. 

\section{Summary and Conclusions}\label{conclude}

In this paper, we first revisit the sensitivity of current and
future CMB experiments to gravitational chirality and to
cosmological birefringence, separately. We show that the WMAP-5
polarization data are not precise enough to provide
any information about gravitational chirality, even for the case
where the tensor-to-scalar ratio is just below the current upper
constraint of $0.22$.  Planck and SPIDER may be able to make a
marginal detection, but only if $r$ and $\Delta \chi$ are both close
to their maximal allowed values.  CMBPol may probe
gravitational chirality over a larger range of the $r$-$\Delta
\chi$ parameter space. As an illustration, the smallest amount of
GW chirality detectable at the 3$\sigma$ level with a
cosmic-variance--limited experiment, if $r$ is just below
$0.22$,
corresponds to about $65\%$ of the GW background being of one
handedness, and $35\%$ of another. In an analogous analysis, we
show that Planck has a $1\sigma$ sensitivity to a CB rotation
angle of about $16'$, while a CV-limited experiment could reach
about 2 $\mu$arcsec.

In the second part of this paper we show that there is no strong
degeneracy between $\Delta \alpha$ and $\Delta \chi$
parameters. In other words, the effects of chiral gravity and CB
can be easily distinguished, provided that the TB/EB
correlations are clearly detected. However, the same results can
be interpreted as to infer that a marginal (e.g., 3$\sigma$)
detection of $\Delta \alpha$ could be due, alternatively, to
gravitational chirality at some level. For example, if CMBPol
were to measure $\Delta \alpha \simeq 15"$ and find $r=0.1$,
that TB/EB detection could alternatively be attributed, with
similar statistical significance, to gravitational chirality
with $\Delta \chi=0.6$.  If, however, the earlier suggestion of
a TB/EB signal corresponding to a rotation angle of $6^\circ$
had held up, it could {\it not} have been attributed to chiral
GWs, as the implied value of $\Delta \chi$ would have been in
the unphysical regime $\Delta\chi \gg1$.

If a parity-violating signal is detected in the CMB and
attributed to CB, it may be possible to test it further with
observations of cosmological radio sources
\cite{Wardle:1997gu}.  Off-diagonal correlations in the CMB may
also provide additional information on CB, if the CB rotation
angle is position dependent \cite{Kamionkowski:2008fp,
Yadav:2009eb,Gluscevic:2009mm}, as suggested in
Refs.~\cite{Pospelov:2008gg}.  A parity-violating signal from
chiral GWs might be distinguished from that due to CB through
direct detection of the gravitational-wave background at shorter
wavelengths \cite{Seto:2008sr}.  Finally, it may be that any signals of
chiral gravity in the CMB may be corroborated, within the
context of specific alternative-gravity theories, by a variety
of other observations and measurements \cite{Alexander:2009tp}.


\begin{acknowledgments}
VG thanks Timothy Morton for useful comments. This work was supported by DoE DE-FG03-92-ER40701 and the Gordon
and Betty Moore Foundation.
\end{acknowledgments}

\section*{APPENDIX A: Power-Spectra Covariance
Matrix}\label{ps_covariance}

Suppose we have obtained multipole coefficients $d_{lm}^{X'}$,
for $X=\{T,E,B\}$ from a full-sky CMB map. Their variance is
given by \cite{Kamionkowski:1996ks}
\beq  
     \left\langle {{{(d_{lm}^X)}^*}d_{l'm'}^{X'}} \right\rangle
     = (|W_l^b{|^2}C_l^{XX'} + w_{XX'}^{ - 1}){\delta
     _{ll'}}{\delta _{mm'}},
\label{dd_mean}
\eeq
where $C_l^{XX'}$ is the power spectrum of the signal, 
$W_l^b \simeq \exp(-l^2\sigma _b^2/2)$ is the window function to
take into account the effects of beam smearing, and
$\sigma_b\equiv \theta_{\mathrm{FWHM}}/\sqrt{8\ln(2)}$ with
$\theta_{\mathrm{FWHM}}$ the beam width.  The $w_{XX}^{-1}$ are
the contributions to the measured power spectra due to
instrumental noise; they are given by
\beq
\begin{gathered}
  w_{TT}^{ - 1} \equiv \frac{{4\pi \sigma _T^2}}
{{{N_{\text{pix}}}}}, \hfill \hspace{0.5cm}
\text{and} \hspace{0.5cm}
  w_{EE}^{ - 1} = w_{BB}^{ - 1} \equiv \frac{{4\pi \sigma _P^2}}
{{{N_{\text{pix}}}}}. \hfill \\ 
\end{gathered} 
\label{ws}
\eeq
Here, $\sigma_T$ and $\sigma_P$ are the pixel noise in
temperature and polarization, respectively, and $N_\text{pix}=4\pi
\theta_{\mathrm{FWHM}}^{-2}$ is the number of pixels.
We assume that the signal is not correlated to the
noise, and that the noise in the polarization is not correlated
to the noise in the temperature; i.e. $w_{ET}^{ - 1} = w_{BT}^{
- 1} = w_{EB}^{ - 1} = 0$. 

The estimators for the power spectrum are then
\beq
     \hat C_l^{XX'} = |W_l^b{|^{ - 2}}\left(
     \sum\limits_{m=-l}^{l}
     {\frac{{{{(d_{lm}^X)}^*}d_{lm}^{X'}}}
     {{(2l + 1)}}}  - w_{XX'}^{ - 1}\right), 
\label{C_estimator}
\eeq
and the power-spectrum covariance matrix is then given by
\beq
\bga
     \Xi ^{{X_1}{X_2},{X_3}{X_4}}_{l} \equiv \left\langle
     {\left(\hat C_l^{{X_1}{X_2}} -
     C_l^{{X_1}{X_2}}\right)\left(\hat C_l^{{X_3}{X_4}} -
     C_l^{{X_3}{X_4}}\right)} \right\rangle \\ 
     = \left\langle {\hat C_l^{{X_1}{X_2}}\hat C_l^{{X_3}{X_4}}}
     \right\rangle - C_l^{{X_1}{X_2}}C_l^{{X_3}{X_4}} \\ 
     = {|W_l^b|^{ - 4}}\Biggl({\sum\limits_{m,m'}^{}
     {\frac{{\left\langle
     {{{(d_{lm}^{{X_1}})}^*}d_{lm}^{{X_2}}{{(d_{lm'}^{{X_3}})}^*}d_{lm'}^{{X_4}}}
     \right\rangle }}{{{{(2l + 1)}^2}}}}}\\
     - {\left\langle
     {{{(d_{lm}^{{X_1}})}^*}d_{l'm'}^{{X_2}}} \right\rangle
     \left\langle {{{(d_{lm}^{{X_3}})}^*}d_{l'm'}^{{X_4}}}
     \right\rangle }\Biggr) \\ 
     =\frac{{|W_l^b{|^{ - 4}}}}
     {{(2l + 1)}}\left({\left\langle
     {{{(d_{lm}^{{X_1}})}^*}d_{l'm'}^{{X_3}}} \right\rangle
     \left\langle {{{(d_{lm}^{{X_2}})}^*}d_{l'm'}^{{X_4}}}
     \right\rangle}\right. \\ 
      \left.{+ \left\langle
      {{{(d_{lm}^{{X_1}})}^*}d_{l'm'}^{{X_4}}} \right\rangle
      \left\langle {{{(d_{lm}^{{X_2}})}^*}d_{l'm'}^{{X_3}}}
      \right\rangle }\right).
\label{covariance_matrix_derivation}
\ega
\eeq
Using Eqs.~(\ref{C_tilde}) and (\ref{dd_mean}) we get
\beq
     \Xi ^{{X_1}{X_2},{X_3}{X_4}}_{l} = \frac{1}
     {{(2l + 1)}}(\tilde C_l^{{X_1}{X_3}}\tilde
     C_l^{{X_2}{X_4}} + \tilde C_l^{{X_1}{X_4}}\tilde
     C_l^{{X_2}{X_3}}),
\label{covariance_matrix_entries}
\eeq
where
\beq
\tilde C_l^{XX'} \equiv C_l^{XX'} + w_{XX'}^{ - 1}|W_l^b{|^{ - 2}}.
\label{C_tilde}
\eeq

To account for partial-sky coverage, we add a factor
$f_{\mathrm{sky}}^{-1}$ (the inverse of the fraction of the sky
used in the analysis) to the right-hand side of 
Eq.~(\ref{covariance_matrix_derivation}).  We also multiply the
factors $w_{XX}^{-1}$ in Eq.~(\ref{ws}) by a factor
$(f_{\mathrm{sky}}^0)^{-1}$ (the inverse of the fraction of the sky
surveyed).

\section*{APPENDIX B: Constraints on Tensor-to-Scalar Ratio}\label{r_constraints}

Here we examine a claim of Ref.~\cite{Contaldi:2008yz} that if
the GW background is chiral, it may be more easily detected
through the TB signal than the BB signal, the reason being that
it may be easier to detect a weak signal (B) by
cross-correlation with a strong one (T) than against itself.

Under the null hypothesis (no GWs), the error with which the
tensor-to-scalar ratio can be measured, from just one power
spectrum A (where A is BB, TB, or EB) is
\beq
     \sigma_r^{-2} = \sum_l \left( \frac{\partial
     C_l^A}{\partial A_t} \right)^2 \left( \Xi_{l,AA}
     \right)^{-1}.
\eeq
Given that the power spectra are simply proportional to $r$,
$(\partial C_l^A/\partial r) \propto C_l^A$.  The relevant
covariance-matrix entries are
\beq
     \Xi_l^{BB} = \frac{2}{2l+1} \left(\tilde C_l^{BB} \right)^2
     =\frac{2}{2l+1} w_{BB}^{-1} (W_l^b)^{-2},
\eeq
\begin{eqnarray}
     \Xi_l^{TB} &=& \frac{1}{2l+1} \left[ \left(\tilde
     C_l^{TB} \right)^2 + \tilde C_l^{TT} \tilde C_l^{BB} \right]
     \nonumber \\
     &=& \frac{1}{2l+1} w_{BB}^{-1} (W_l^b)^{-2} \left[ C_l^{TT,s} +
     w_{TT}^{-1} (W_l^b)^{-2} \right],\nonumber \\
\end{eqnarray}
(and similarly for EB, with T$\to$E) where we have employed the
null hypothesis in the second equality in each of these equations.

Given that $w_{TT}^{-1} \ll C_l^{TT}$ already from current data
for the low $l$ at
which the GW signal arises, we can set $w_{TT}^{-1}=0$.
Moreover, $C_l^{TB}\sim \beta (C_l^{BB} C_l^{TB})^{1/2}$, with
$\beta\sim 0.1$.  As a result, while the summand for $\sigma_r^{-2}$
from BB is $\sim (C_l^{BB})^2/w_{BB}^{-2}$, that from TB is
$\sim C_l^{BB}/w_{BB}^{-1}$.  Thus, in the limit of sufficiently high
signal-to-noise, $w_{BB}^{-1}\to  0$, the BB signal provides a
better probe (smaller $\sigma_r$).  In other words, the value of the
cross-correlation with T is ultimately limited by cosmic
variance (as is also the cross-correlation with E), while the BB
sensitivity improves without limit as the
instrumental noise is reduced.  (The importance of TB is also
weakened slightly given that $C_l^{TB}< [C_l^{TT}
C_l^{BB}]^{1/2}$.)  It is true that in the 
opposite limit, where $w_{BB}^{-1}$ is large, TB is more
sensitive to GWs (with $\Delta \chi=1$) than TT.  However, this limit
is only of academic interest, as it is encompasses the regime of
$r$ that is already ruled out by temperature measurements.

To make these arguments more quantitatively precise, we have
evaluated $\sigma_r$ for BB, TB, and EB (for $\Delta\chi=1$) for
WMAP-5, SPIDER, Planck, and CMBPol; the results are shown in
Table \ref{rs}.  We see that the sensitivity to GWs with future
experiments will come primarily from BB, with only marginal
improvement from TB.  While the TB sensitivity of WMAP-5 is better
than that from BB, the smallest $r$ detectable with either is
already larger than the upper limit from TT.

The bottom line:  While TB may improve the sensitivity to a
chiral-GW background, it does so only marginally, with most of
the sensitivity due primarily to BB (see also Ref.~\cite{Saito:2007kt}).

\begin{center}
\begin{table}[htbp]
    \begin{tabular}{ | l | l | l | l | p{5cm} |}
    \hline
    Instrument & from BB & from TB & from EB\\ \hline
    WMAP-5 & 0.68 & 0.37 & 3.03 \\ \hline
	 SPIDER & 0.011 & 0.051 & 0.20 \\ \hline
    Planck & 0.026 & 0.071 & 0.30 \\ \hline
    CMBPol & 1.57$\times 10^{-5}$ & 0.0018 & 0.0062 \\
    \hline
    \end{tabular}
\caption{The error $\sigma_r$ on the tensor-to-scalar ratio
    for a chiral GW background with $\Delta\chi=1$ from BB, TB,
    and EB for several CMB experiments.\label{rs}}
\end{table}
\end{center}


\end{document}